# Through Silicon Vias as Enablers for 3D Systems

Erik Jung*, A. Ostmann*, P. Ramm**, J. Wolf*, M. Toepper*, M. Wiemer***
Fraunhofer IZM
*Berlin/**Munich/**Chemnitz
Contact: Erik.jung@izm.fraunhofer.de

*Abstract*-This special session on 3D TSV´s will highlight some of the fabrication processes and used technologies to create vias from the frontside of an active circuit to its backside and potential implementation solutions to form complex systems leveraging these novel possibilities. General techniques for via formation are discussed as well as advanced integration solutions leveraging the power of 3D TSV´s.

## I. Introduction

For decades, ideas to bring contacts from the highly populated frontside of electronic circuitry has been the holy grail of integration techniques. Since the early 1990´s, technical progress has enabled semiconductor engineers to continuously work towards this goal /1,2,3,4/. Processes that allowed the formation of holes through the bulk silicon, allowed to thin down the bulk substrate to minimize the overall process time, and the development of deposition processes to insulate and fill the created holes, forming electrically conductive vias are enablers for today´s 3D TSV´s.

It took until recently, to bring all these processes and the associated quality assurance to a point, where synergy really allows to define processes capable of being implemented in volume manufacturing. This paper provides an overview on established techniques for:
- thinning of wafers towards 150…5µm remaining substrate thickness
- creating well defined holes through the bulk material
- insulating the semiconductor material and
- forming a conductive interconnect between the pads on front and backside of the device

In addition, alternatives to these process sequences as well as methods to integrate such devices in a complex system are reviewed.

## II. Processes for Via Formation

Generally, the concepts to route contacts from the front to the backside can be classified in two categories:
a) routing of multiple contacts towards and through large via holes with thin film lines and re-routing them on the backside to their final position
b) creating through-holes underneath or in the pad to be routed, insulating and filling the via
c) creating through-holes in the pad to be routed, insulating and filling the via.

Concepts like a) (**Figure 1** a and b) have been used to create lidding structures for MEMS and RF devices as well as CSP type of integrated systems for MEMS components /5,6/. They do not fall in the current focus of research interest of complex microelectronic circuit manufacturers, as they do not provide the required via density for modern chips, e.g. coming with sub 100µm contact pitches

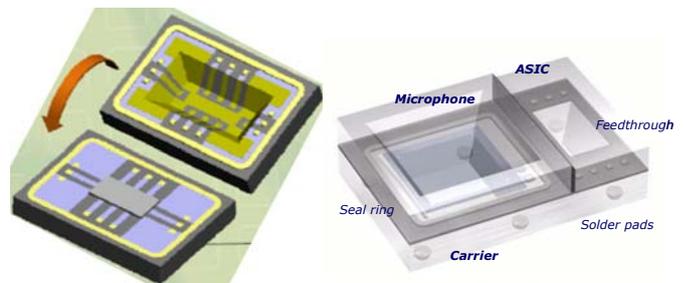

Figure 1: Routed via concepts through KOH etched holes for a) RF (courtesy Hymite) and b) MEMS systems (courtesy Sonion MEMS)

Concepts like b) & c), depicted in **Figure 2** a and b are more suitable to the requirements of high density circuits, routing a couple of hundred contacts to the backside of a single chip.

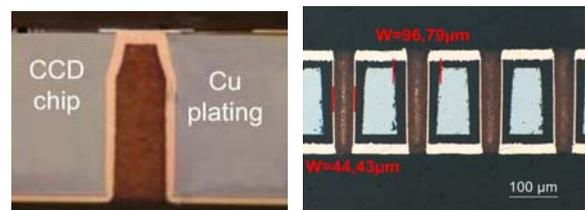

Figure 2: Blind TSV via with side wall copper fill (courtesy University of Arkansas), laser drilled through via with side wall fill based on laminate dielectric /[7]/

As silicon wafers coming out of a fab have thicknesses ranging from 500 with 100mm wafers to 800µm for 300mm wafers, all currently available processes to create holes are hard pressed to realize fast, high quality and precise micro holes. With the advent of mainstream thinning technology, these finalized wafers are backgrinded to 150µm remaining silicon and less. Table I shows the target thickness and the suggested processes to achieve a high quality surface.

TABLE I - Processes for Wafer Thinning

| Target thickness | Process | Comment |
|---|---|---|
| ~300um | Coarse Backgrinding | Stress relief etch suggested |
| ~100um | Fine Backgrinding, stress relief etch | After Coarse Backgrinding, Stress relief etch using wet etching or Atmospheric Downstream Plasma |
| ~50µm | Fine Backgrnding, etching | Similar like above, better parameter control, slower process |
| ~10µm | Wet etching or ADP etching | Backside crack removal becomes mandatory to increase devices stability |
| 10µm | Wet etching + CMP | Photo receptors start to be influenced by missing substrate |
| ~2µm | Intermediate layer release or CMP | Removal of nearly all non-CMOS-functional silicon |

For the subsequent processes, typically the range of 150…10µm is targeted, thereby allowing to use of today´s standard processes of backgrinding, etching and CMP /[1]/.

At these remaining silicon thickness, hole drilling techniques as of Table II become economical even with the required precisions.

TABLE II
Hole Drilling Techniques

|  | Technique | | |
|---|---|---|---|
|  | Wet Etching | Plasma (DRIE) Etching | Laser drilling |
| Hole fabrication speed | 1..11µm/min | Up to 50µm/min | 2400 vias/s |
| Position precision | Mask defined ++ | Mask defined ++ | Conveyer system: several µm |
| Aspect ratio | 1:1…60 | 1:80 | 1:7 |
| Precision | Sub um | Sub µm | ~10µm |
| Quality of hole | excellent | Good (scallops) | Very good |
| Example figure | Figure 3a | Figure 3b | Figure 3c |

Sources: /8, 9, 10/

---

[1] Processes may target the formation of TSV´s before or after the actual processing of the electronic circuit. Respective denominators refer to FEOL and BEOL via formation.

While the etching processes can be tailored to stop on the pad and under pad metallization layers, this is more critical with the laser. Here, before the thinning process, usually some additional µm of a compatible metal are deposited on top of the existing pads. Drilling into or even through these enables also a larger interconnect annulus.

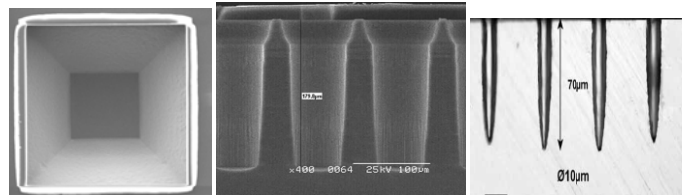

Figure 3: a) KOH etched silicon @ 54.7°;
b) DRIE etched via (courtesy of ALCATEL);
c) Laser Drilled Vias (courtesy of XSIL)

Silicon as semiconductor material requires an insulative cover of the hole´s sidewalls. For this, low temperature PECVD oxide deposition or organic insulators like PI or BCB are used. Finally, CVD (tungsten) or electro(less) deposition of copper is used to fill the vias (**Figure 4** a and b).

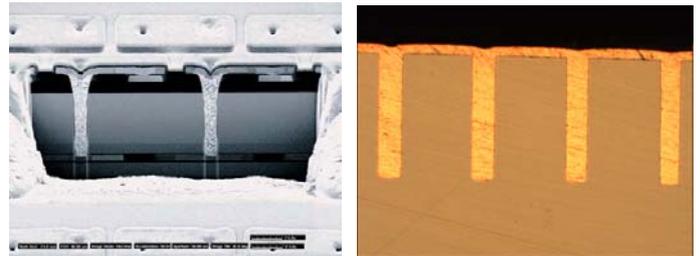

Figure 4: Solid filled vias a)W-CVD (courtesy IZM),
b) Cu-ED (courtesy NEXX)

A variation of this filling is just to cover the sidewalls and use a high step coverage process to route thus created contacts to the final pads /11 / (**Figure 5**).

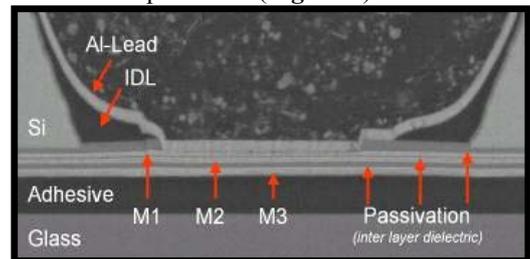

Figure 5: Thin film via contact (courtesy SCHOTT)

III. SYSTEM INTEGRATION CONCEPTS

After the described process sequence, a device with backside contacts is available. For the system integration concepts, this is just another technology in the tool box to create highly integrated complex functions. From the system aspect, both sides –front and backside- are accessible for interconnect with other devices. TSV`s currently target mainly memory chips like FLASH and DRAM, as these have identical sizes, identical pad-arrangements and have a high wafer yield, making them ideal for a wafer-to-wafer integration (**Figure 6**).

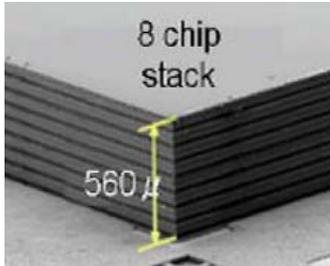

Figure 6: 3D Chip Stack using TSV´s
(courtesy Yole)

However, system integration does not stop there. The tool boxes allow also to integrate non-identical devices (hetero-system integration) using state of the art assembly and packaging techniques.

Redistribution processes in combination with thin chips allow multiple layers of circuitry to be integrated on one chip-substrate (Figure 7) or as a complex stack (Figure 8).

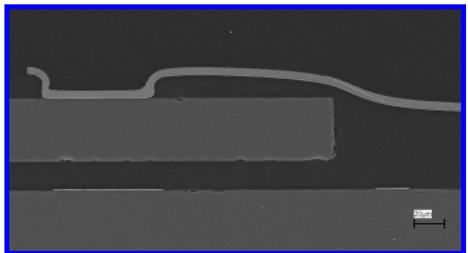

Figure 7: Thin chip integration on chip

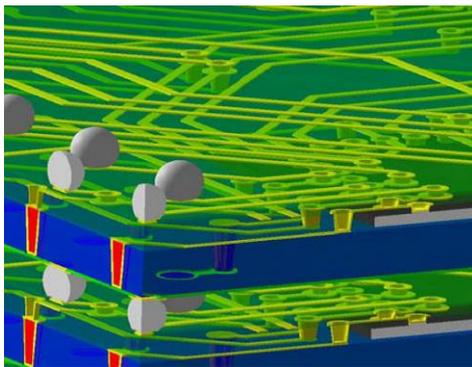

Figure 8: Top/bottom routed set of chips for
3D complex system integration
(courtesy iNEMI, IZM) /12/

Optical systems have been demonstrated to benefit from 3D TSV´s (Figure 9 and Figure 10) /13/ similar like MEMS devices (e.g. for mirror devices).

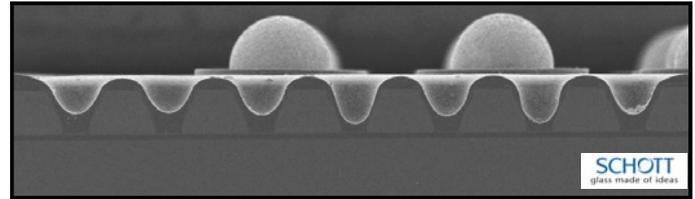

Figure 9: 3D TSV used for a camera chip in CSP format
(courtesy SCHOTT)

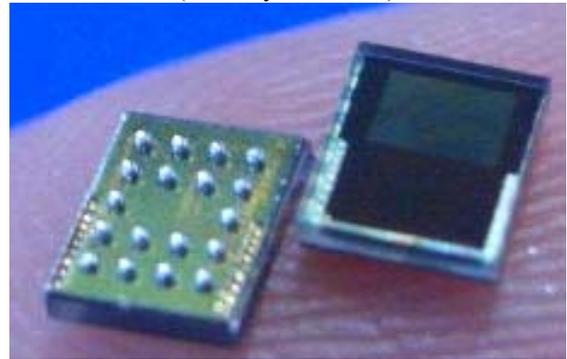

Figure 10: True wafer level CSP opto package
(courtesy SCHOTT)

With optical systems, the combination of wafer-to-wafer integration to merge the optical function of a lens layer with the sensor chips, a hybrid integration on the connected backside allows to build extremely small and lightweight fully integrated cameras with enhanced functionality (Figure 11).

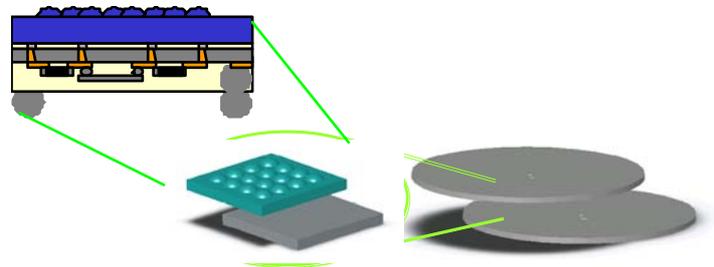

Figure 11: Hybrid integration of additional circuitry on the backside of a TSV enabled camera system

MEMS integrated systems can benefit from short signal paths and ultrasmall integration /14/.

In addition, hybrid integration targets directly two of the biggest issues with the concept of wafer-to-wafer interconnects: Chip size and chip yield.

As mentioned, with memory chips running at some 95% yield and identical dimensions, microprocessors, MEMS and CMOS optical sensors have a significantly lower yield (down to ~30%) and are –by nature- of different size. A wafer-to-wafer interconnect would require a huge sacrifice of silicon for the smaller chips and a multiple yield loss. Hybrid integration foregoes this problem by enabling the use of "known good dies".

Hybrid integration is conducted using advanced pick and place equipment and the respective interconnect processes like flip chip or thin film interconnect for chip-first approaches (e.g. shown in Figure 7). The pick&place process enables also the use of small SMD devices, indispensable in forming today´s integrated systems.

## IV. CONCLUSION

3D TSV´s have come a long way since their first conception. While the general packaging advancement has still an edge w.r.t. manufacturing, it is quite obvious that future advancement in 3D TSV as well as associated integration techniques will enable integrated system densities and new products that have not been possible until now.

## V. ACKNOWLEDGEMENT

The authors would like to thank the collaborators from SCHOTT, EMC-3D (www.emc3d.org) and iNEMI (www.inemi.org) for the support and images supplied